\documentstyle[11pt,gh2001-asp,twoside,epsf]{article}
\input psfig.sty
\def\edcomment#1{\iffalse\marginpar{\raggedright\sl#1\/}\else\relax\fi}
\reversemarginpar

\begin{document}
\title{Density waves in the inner parts of disk galaxies}
 \author{Eric Emsellem}
\affil{Centre de Recherche Astronomique de Lyon, 9 av. Charles Andr\'e, 69561 Saint Genis
	Laval Cedex, France}

\begin{abstract}
Density waves in the central kpc of galaxies, taking the form of spirals,
bars and/or lopsided density distributions, are potential actors of the
redistribution of angular momentum. They may thus play an important role
in the overall evolution of the central structures, not mentioning the
possible link with the active/non-active central mass concentration.
I present here some evidence for the presence of such structures, and discuss their
importance in the context of dynamical evolution.
\end{abstract}

\section{Introduction}

Spirals, bars, warps and lopsidedness are ubiquitous at large scales in disk galaxies
(e.g. Sancisi 1976; Kamphuis et al. 1991). These structures are 
usually understood as density waves propagating in the stellar and/or gaseous component.
In this paper, I emphasize the case for density waves in the central
kpc of disc galaxies, and illustrate their presence and role with a
few examples.

\subsection{Role of density waves}

Density waves such as bars, spirals and lopsided density distributions have been
recognised as potential actors in the secular evolution of galaxies,
partly because they are non-axisymmetric perturbations of the potential.
Gas flows within a bar (Athanassoula 1992) 
are very illustrative of how such structures can redistribute angular momentum:
gas can be transported e.g. inwards (and angular momentum outwards) 
via driven trailing spirals,
strong shocks can occur and rings may form (usually near specific resonances).
All these correspond to observed signatures of bars. 
Density waves also tend to heat the stellar system,
therefore acting against subsequent gravitational instabilities.
Small versions of the large-scale
waves observed in disk galaxies have been invoked as responsible
for the fueling of the central 10s of parsecs, and hypothetically
the central AGN and/or starburst (Schlosman, Frank \& Begelman 1989; Regan \& Mulchaey 1999). 
But are these inner bars/spirals modes similar to their large-scale parents?

\subsection{Towards the centre}

A dynamical probe which would travel towards the centre of a galaxy would 
see a number of changes occuring:
spatial sizes are getting smaller and time-scales shorter (a useful
number to keep in mind is: $10^6\,$yr corresponds to $\sim 100$~pc at 100~km.s$^{-1}$), 
the spheroidal component starts to significantly contribute to the potential, 
and eventually, at a scale of a few parcsecs, 
the central cusp and/or supermassive black hole may dominate
the potential. Within the central kpc, an inner cold component (e.g. a disk) 
may be less self-gravitating, and tend to evolve more rapidly. These changes
obviously affect the growth and propagation of density waves in the inner parts
of galaxies. As observers better focus on the central kpc of disk
galaxies (with the help of an improved spatial resolution),
structures such as inner spirals, bars and lopsided density distribution
are revealed (Erwin et al. 2001; Regan \& Mulchaey 1999), and are indeed usually
weaker than their large-scale versions. Before these are 
confirmed as true density waves however, their kinematics should be studied in
more detail. After a brief reminder on resonances, I will provide 
a few examples of such observations.

\subsection{Epicycle approximation}

Circular orbits can be used as zeroth order approximations for orbits in the 
equatorial plane of a thin disk. The first order terms 
can easily be evaluated by linearising the equations of motion in that plane:
a retrograde epicycle motion is added on the circular orbit.
This epicycle is the combination of radial and azimuthal
harmonic oscillations sharing the same frequency $\kappa$,
which only depends on the first and second radial derivatives of the potential
(see Dehnen 1999 for an alternative view on Lindblad's epicycle theory).
The shape of the (first order) orbits at a certain radius will then depend on the
ratio $\kappa / \Omega$ where $\Omega$ is the circular frequency: when
this ratio is an integer $m$, the orbits are periodic (closing after $m$ epicycles).
With the addition of a density wave with a pattern speed $\Omega_p$,
the potential stays constant only in a frame rotating with the wave.
In this rotating frame, the important quantity which decides on the shape of the orbit
then becomes $\kappa / \left(\Omega - \Omega_p\right)$.
The resonances are thus located at radii where this ratio takes integer ($m$) values:
$ \Omega_p = \Omega + \kappa / m $.
This is where the disc potential (with its natural frequency $\kappa$) 
and the wave (of angular frequency $\Omega_p$) may interact. 
The most important are the Lindblad Resonances (LR; the Inner LR
or ILR for $m=2$, and the Outer LR, or OLR for $m=-2$) and the Corotation
Resonance (CR) where $\Omega = \Omega_p$ ($m \rightarrow \infty$). At the resonances,
the (linearised) orbits are closed in the frame rotating with the wave, 
thus defining the different families of orbits from which the skeleton
of the system is built.

\section{Inner disks and resonances}

Seifert \& Scorza (1996) found that a significant fraction ($\sim 60$\%) of 
S0 galaxies contain two embedded disks: an outer (main) disk with an 
inner cut-off, and an inner disk. This was confirmed by high resolution HST images
which revealed very thin disks in the central kpc of early-type disk galaxies.
In the case of the Sombrero galaxy (M~104)\index{object, M 104}, the double disk structure was claimed
to originate from bar driven secular evolution, the inner disk
being formed within the ILR of the bar (Emsellem 1995). 
For early-type disk galaxies, such a scenario seems to be supported
by several studies (Scorza \& van den Bosch 1998; Baggett, Baggett \& Anderson 1998).
If the disk hierarchy reflects the existence of resonances,
these are also often emphasized by gas and stellar rings 
(at the OLR and Ultra Harmonic Resonance in the case of M~104). 
Rings are indeed very common signatures of bars (see Combes 2001, 
Buta \& Combes 2000).  

A remarkable example, the nearly edge-on
galaxy NGC~4570\index{object, NGC 4570}, was reported by  van den Bosch \& Emsellem (1998).
This galaxy seems to possess two rings which correspond to the
ILR and Ultra Harmonic Resonance associated with a tumbling potential.
The thin inner disc, with a radius of about 100~pc, is located inside
the ILR. The stellar $U-V$ colour distribution
also seems to reflect the location of the presumed resonances:
a rather flat colour gradient along the major axis outside the CR,
a mild gradient between the CR and the ILR, and a strong one inside
the ILR (Fig.~1). This closely follows the predictions of the simulations
performed by Friedli, Benz \& Kennicutt (1994) who studied the effect
of a bar on the abundance profiles (model d of their Fig.~1). In this scenario,
the inner disk is the result of gas accretion by the large-scale bar.
\begin{figure}[ht]
\centerline{\psfig{figure=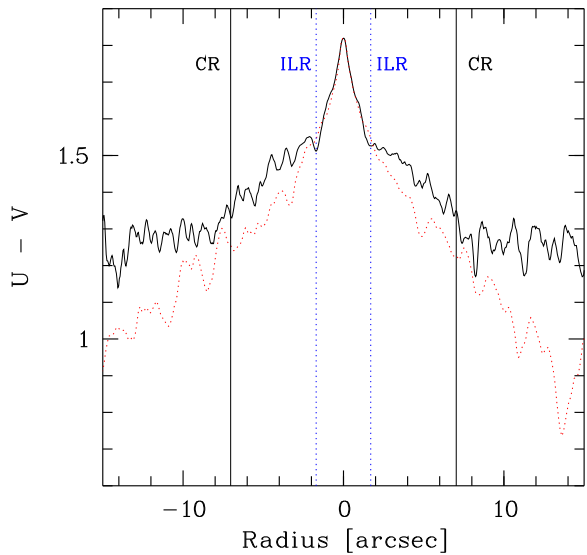,width=8cm}}
\caption{$U - V$ major (solid line) and minor (dotted line)
axis profiles of NGC~4570 (WFPC2). The expected locations of
the resonances linked with a bar are indicated (van den Bosch \& Emsellem 1998).}
\end{figure}

\section{Inner bars}

In some cases, the inner disk can decouple from the outer part,
and a secondary bar can form (Shlosman, Frank, \& Begelman 1989).
Double barred galaxies are now well studied (photometrically), 
and preliminary statistics indicate that at least 25\% of 
barred galaxies host an inner bar
(Erwin et al. 2001). We know that primary bars are efficient at 
concentrating the gas in the central region (Sakamoto et al. 1999). 
Secondary bars could then take the batton and fuel the central
100~pc, although no strong dynamical evidence has been presented yet.
In this context, we recently obtained the stellar kinematics of
4 active galaxies (central starburst and/or Seyfert), 3 of which
are double barred. In these 3 galaxies, the stellar velocity dispersion
profiles exhibit a significant but unusual drop at the centre: evidence
for the presence of a cold component (Emsellem et al. 2001). We interpreted this 
as recent gas accretion triggered by the inner bar, and
subsequent formation of an inner stellar disk (Greusard et al. 2002, 
Wozniak et al. 2002). The sample critically needs to be extended
to understand if there is indeed a link between the central activity
and the inner bar. 

Studies on samples of disk galaxies 
tend to give slightly different answers (Ho, Filippenko \& Sargent 1999;
Knapen et al. 2000), although the recent study by Knapen et al. 
indicate a higher fraction of bars in Seyferts ($79 \pm 7.5$\%) than
in non active galaxies ($59 \pm 9$\%). It seems that
even if there is a tendency for Seyferts to be more barred than 
non active galaxies, it is a {\em weak trend}. As emphasized by Combes (2001),
this is not at all surprising, since there are very different
timescales involved: gas accumulation driven by the bar,
secondary bar formation, star formation, AGN duty cycle, dynamical evolution
including the possible destruction of the bar due to central mass accumulation... 
\begin{figure}[ht] 
\centerline{\psfig{figure=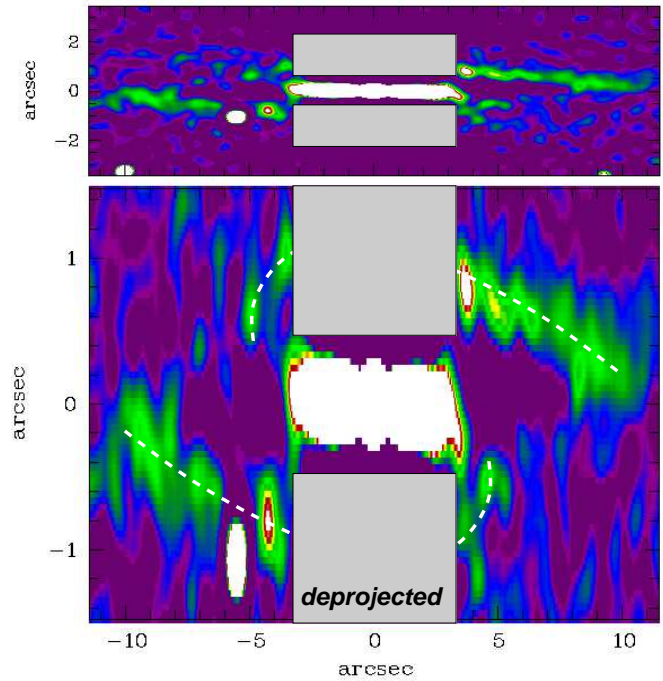,width=9cm}}
\caption{Top panel: unsharp masking of a WFPC2/F555W ($V$ band) image of NGC~3115.
Bottom panel: same image but deprojected using 
an inclination of $86\deg$ (Emsellem et al. 1999).
An hypothetic sketch for a two-arm spiral is superimposed on the deprojected image.
High frequency structures could not be derived in the grey areas due
to the presence of the bright inner disk.}
\end{figure}

\section{Inner spirals}

A recent addition to the usual suspects is the loss of angular momentum
by inner trailing spirals. Such structures are often observed in spiral
galaxies where resolution permits (Regan \& Mulchaey 1999). 
These are low amplitudes modes however, which does not favour an significant central fueling rate,
and these may just be driven by a mildly triaxial tumbling potential.
Inner spirals however require a cuspy mass distribution in order to be trailing
and drive gas inwards to the nucleus. One strong inner gas spiral has 
been observed in the early-type galaxy NGC~2974\index{object, NGC 2974}
(Emsellem \& Goudfrooij 2002). Strong streaming motions were measured using
integral field spectroscopy, and the overall gas kinematics can be qualitatively
reproduced by a simple two-arm spiral model. The spiral ends outside its ILR,
which implies that these do not correspond to
the bar driven acoustic waves described by Englmaier \& Shlosman
(1999). The inner gaseous spiral in NGC~2974 is a rather unique case, 
although a larger observing campaign using the same techniques is on-going.

In the S0 galaxy NGC~3115\index{object, NGC 3115}, we have detected a low contrast stellar spiral
which seems to be directly connected to the inner thin disk 
(Fig.~2). Taking into account the presence of a supermassive black 
hole of $\sim 6.5\,10^8$~M$_{\odot}$ at the centre of NGC~3115 
(Emsellem et al. 1999), it is probable that any strong bar has been significantly
weakened, but the spiral could still be driven by a weak residual triaxiality
in the potential. If the present morphology of NGC~3115 is indeed the result 
of bar driven evolution, we might be able to find clear signatures 
by conducting a coupled study of the stellar populations and dynamics.

\section{$m=1$ modes}

Lopsided distribution have long been ignored, but are now more
often studied theoretically, and looked for in galactic central 
regions (see Combes 2001 and references therein). The short dynamical
timescales in the central kpc of galaxies seem to favour the hypothesis
of strongly lopsided distribution to be true $m=1$ modes. These modes
can often be superimposed on $m=2$ bar modes. One illustration
of this is the double-barred galaxy NGC~3504\index{object, NGC 3504}. The $K$ band adaptive
optics image (courtesy of F. Combes) shows a very strongly asymmetric
light distribution. At visible wavelengths, a one arm spiral
is revealed by the HST/WFPC2 F606W image. We have observed this galaxy
using the integral field spectrograph {\tt OASIS} at CFHT, and determined
the two-dimensional stellar and gas distribution and kinematics.
The spiral arm is strongly emphasized in the emission line maps,
except in the higher density [OI] line, the distribution of
which seems to bridge the inner end of the spiral to the nucleus (Fig.~3).
In fact, detailed examination of the central {\tt OASIS} spectra
reveals the presence of a blueshifted wing in all emission lines
including the forbidden ones. We interpret this as evidence for
nuclear inflow, possibly driven by the $m=1$ mode.
\begin{figure}
\centerline{\psfig{figure=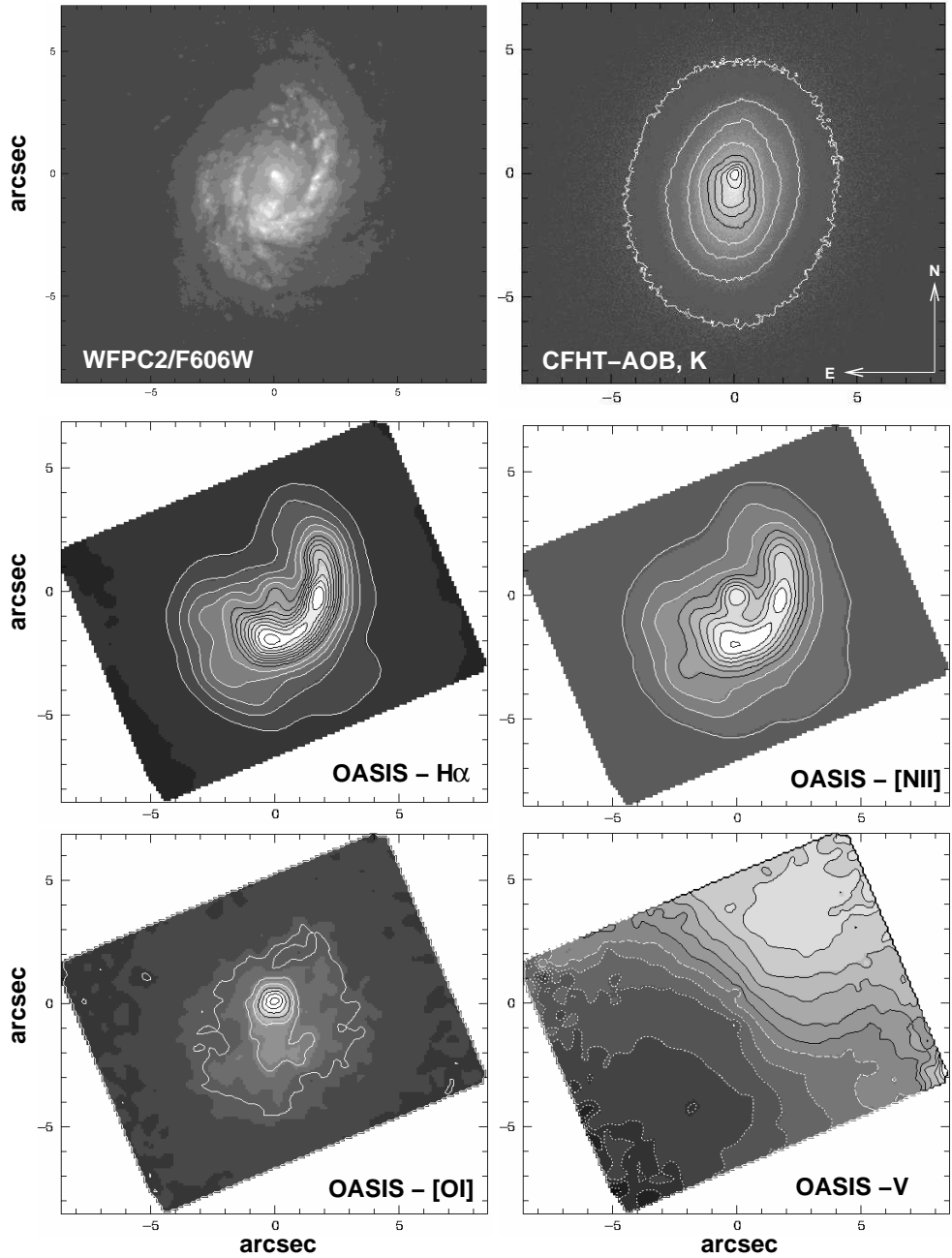,width=13cm}}
\caption{The central 800 pc of NGC~3504. Top left panel: HST/WFPC2 F606W image.
Top right panel: PUEO/CFHT $K$ band image (FWHM $\sim 0\farcs15$).
All other panels: OASIS/CFHT maps of H$\alpha$ (middle left), [NII]$\lambda$6583
(middle right), [OI]$\lambda$6300 (bottom left) and mean velocity (bottom left).
The steps and first contours for the OASIS maps are (12,12), (5,8) and (0.17,0.8) for
the H$\alpha$, [NII] and [OI] maps respectively (units 
of $10^{-15}$~erg/s/cm$^2$/arcsec$^2$).
The isovelocity step is 20 km/s with the systemic velocity being taken at the centre.}
\end{figure}

\subsection{Keplerian modes}

In a keplerian potential, $\Omega = \kappa$, which implies that the precession
rate of the $m=1$ orbits is zero at all radii. A set of such orbits could then
keep their alignement with time. In a marginally self-gravitating disk, gravity could
act to compensate for the non zero precession rate. This seems to be the case
for the double nucleus of M~31, in which the offcentred peak could be explained
by the orbit crowding (see Bacon et al. 2001, and references therein).
This sets a limit on the relative mass of the nuclear disk, which should
be between 20 and 40\% of the central dark mass. The nuclear disk
of M~31 could have been formed relatively long ago by gas accretion
triggered by a tumbling potential. 
In fact, several pieces of evidence
support this hypothesis: M~31\index{object, M 31} is very probably a barred galaxy (Berman 2001),
and the inner dust structures are consistent with being trailing spiral arms
nearly reaching the nucleus (Fig.~4). There is finally a hint of a recent but limited
episode of star formation at the very centre of the nucleus (Fig.~5). The $m=1$
mode could then be the result of either a natural instability, or 
a relatively recent gas cloud infall.
\begin{figure}
\centerline{\psfig{figure=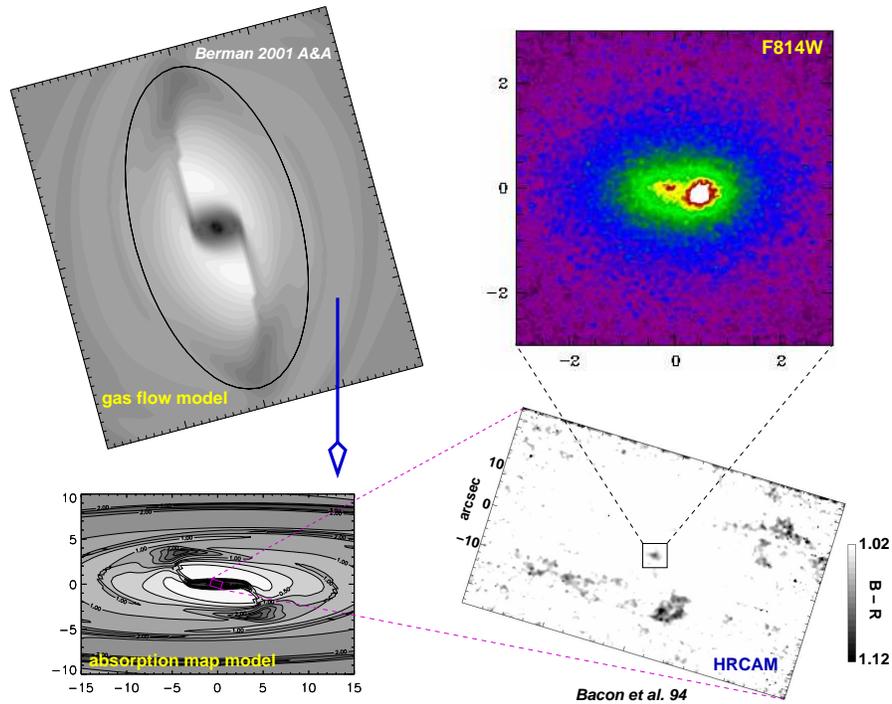,angle=-90,width=15cm}}
\caption{Zooming from the large-scale bar of M~31
to the central spiral-like dust lanes, and the 10~pc double nucleus.}
\end{figure}
\begin{figure}
\centerline{\psfig{figure=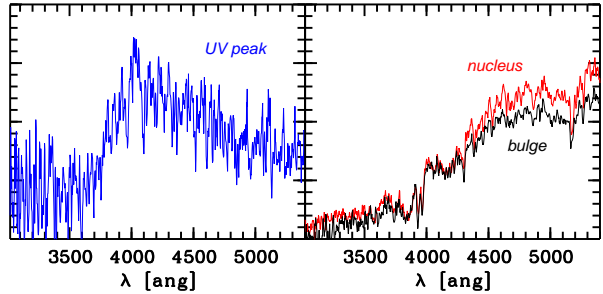,width=8cm}}
\caption{HST/STIS spectra of M~31 \index{object, M 31}(not corrected for extinction). 
Left: UV peak contribution. Right: nucleus and bulge spectra.}
\end{figure}

\section{Conclusions}

The short timescales associated to the physical processes
occuring in the central regions of galaxies do not help
our understanding of their morphology and dynamics.
We should then view them as continuously
evolving systems involving several recurrent interlinked 
and non simultaneous processes. Density waves are certainly
playing an important role in this game. If inner bars and spirals
are already on the priority list of observers and theoreticians,
then $m=1$ modes must now be seriously
considered as potential players on the evolutionary stage of galactic centres.

\end{document}